\documentstyle[aps,prb,epsfig,twocolumn]{revtex}

\begin{document}

\bibliographystyle{prsty}

\draft

\wideabs{
\title{Possible Laws for Artificial Life Evolution}
\author{ P. Ao }
\address{ Departments of Mechanical Engineering and Physics, 
          University of Washington, Seattle, WA 98195, USA }
\date{ March 8, 2004 }

\maketitle

\begin{abstract}
   Motivated by a recent article on open problems in artificial life, here I postulate three laws which form a mathematical framework to describe artificial life evolutionary dynamics. They are based on a continuous approximation of population dynamics. Four dynamical elements are required in this formulation: ascendant matrix, transverse matrix, fitness function, and the stochastic drive. The first law states that in the absence of stochastic drive the artificial life always seeks for a local fitness attractor and stay there. It gives the reference point to discuss the general evolutionary dynamics. The second law is explicitly expressed in a unique form of stochastic differential equation with all four dynamical elements. The third law defines the relationship between the focused level of description to its lower and higher ones, and also defines the dichotomy of deterministic and stochastic drives.  These laws provide a coherence framework to discuss several current problems, such as emergency and stability. In particular, two quantities are emphasized: the fitness function as the standard for selection and the stochasticity as the source of creativity.
Those three laws may appear almost self-evident from a statistical physics point of view. However, their equivalent to a most conventional approach for evolutionary dynamics is shown for the first time by the present author, to the best of his knowledge. The computational advantage of the present formulation in the study of artificial life evolution is also discussed.     
\end{abstract}

}


\section{ Introduction }

In a provocative article jointly published by Bedau {\it et al.} \cite{open}, fourteen open problems are formulated to indicate long term research directions in Artificial Life research. They consist of problems to address three types of general concerns. The first type is to explore the relationship between real biological life and its material constituents, that the relationship between biological and physical sciences. The second type is to find the quantitative descriptions of the conditions and potentials for living systems. 
The third type is on consciousness and culture in the context of living systems. 
Those problems provide a very comprehensive framework for forming research programs and they echo fundamental questions raised previously \cite{farmer1}.
Specifically, the 8th problem has been stated as follows: Create a formal framework for synthesizing dynamical hierarchies at all scales. 

The present article is an attempt to address the 8th open question in Ref.[1]. It may also be relevant to address several other relevant quantitative open questions. Previous attempts have been made \cite{rasmussen} to address the 8th question. Furthermore, there have been extensive literature on it \cite{farmer1,al2}. Such a fundamental question has also been raised and the constraints for possible solutions have been discussed \cite{simon,etxeberria,jones}. 
Given above consideration, the author believes that the present approach differs from those previously in the following aspects: 
It is more abstract, more quantitative, and more hierarchical.

The present mathematical approach is based on the continuous approximation which treats the populations as continuous variables. This approximation has been well studied in physical sciences \cite{vankampen} as well as in artificial life \cite{adami}. 
Apart from a discussion on the connection of present approach to the genetic algorithm, this approximation will not be elaborated further. 
This implies that the equations to be discussed are of differential equation type. To be more precise, I will postulate three laws for evolution, and the most important law, the second law, will be expressed in a unique form of stochastic differential equation.
The critical component for the broad scope application of those three laws lies in the explicit demonstration of the equivalence between those three laws and a most general conventional approach to the evolutionary dynamics. 

Two quantities are emphasized in the present formulation. The first one is the fitness landscape. A scalar nonlinear function, fitness function, is introduced to describe it. Though its usage has a long history, but the fitness function dose not appear to occupy a prominent position in current theoretical formulations \cite{farmer2,hofbauer}, and its meaning in evolution has been frequently questioned \cite{farmer1}. This might be partially due to the well-known failure to find potential in neural network computation for general asymmetric nets, such as in cases of limit cycles \cite{hertz}. I will present an explicit demonstration of how to construct a potential, the fitness function, below. I will argue that this fitness function is needed in any evolution, as long as there is a selection: Fitness function is the standard for selection.    

The second quantity is the stochasticity. It is the source for innovation, creation and surprise, and for breaking down the curse of determinism. 
I will show that in its absence the evolution will be stuck to local attractor and will not be able to move to another attractor, regardless it would be better or worse. Its importance has been indeed emphasized in the study of Artificial Life \cite{farmer2,channon}. Nevertheless, in some approaches to evolutionary dynamics, its importance seems not to be fully appreciated \cite{hofbauer}. In the present formulation, the importance of stochasticity will be formulated in a particular transparent manner: It is the engine of evolution.

I organize the rest of article as follows. In section II I postulate and discuss the three laws of evolution. The most important law is the second. In section III the connection of the postulated three laws to previous formulations is discussed. This section lays down the foundation for the generality of the three laws proposed in section II. An explicit computation of this connection is demonstrated in section IV. I also discuss the compatibility of present formulation with an interesting speciation model. The implications of present formulation were discussed in section V. In particular, its computational advantage is pointed out. I conclude in section VI. 

\section{ Laws for Artificial Life Evolution }

In this section I postulate and discuss three laws of evolution. 
They form a quantitative mathematical framework for evolutionary dynamics, with four dynamical elements. Then I discuss a fundamental theorem of evolution.

\subsection{ First Law }

The first law is a statement when there is no stochastic drive in the evolution. This law connects the semi-positive definite symmetric ascendant matrix $A$, anti-symmetric transverse matrix $T$, and the scalar function of fitness function $\psi$ in the following mathematical equation:
\begin{equation}
  [ A({\bf q}_t, t) + T ({\bf q}_t, t) ] \dot{\bf q}_t 
     = \nabla \psi ({\bf q}_t, t)  \; .
\end{equation}
Here I have considered an $n$ component artificial life system. 
The $n$ components may be the species \cite{maynardsmith}, or the traits to describe the speciation \cite{stewart}, or any quantities required to specify the system. The value of $j^{th}$ component is denoted by $q_j$. The $n$ dimensional vector ${\bf q}^{\tau} = (q_1, q_2, ... , q_n)$ is the state variable of the system. Here the superscript $\tau$ denotes the transpose. The dynamics of state variable is described by its speed $ \dot{\bf q}_t \equiv 
d {\bf q}_t /d t $ moving in the state space.
To ensure the independence of the dynamics of each component, I assume $\det [A({\bf q}, t) + T({\bf q}, t)]\neq 0$.
Here the subscript $t$ denotes that the state variable is a function of time and $\nabla$ is the gradient operator in the state space.

The anti-symmetric matrix permits 'no change': 
$ \dot{\bf q}^{\tau}_t T({\bf q}_t, t) \dot{\bf q}_t = 0 $, 
therefore conservative. Dynamically it will not change the fitness.
A manifestation of the transverse dynamics is the oscillatory behavior.   
It is required by the consistency of present mathematical formulation to account for the asymmetric nature of general evolutionary dynamics.
The physical analogous may be the mass in Newtonian dynamics, where the oscillation of a pendulum is guaranteed by the conservation of energy: sum of kinetic and potential energies, or, the vector potential in electrodynamics needed for electromagnetic waves.  

Because of the ascendant matrix $A$ is non-negative, the system will approach the nearby attractor determined by its initial condition, and stay there forever. Specifically, because $ \dot{ \bf q}^{\tau}_t A({\bf q}_t, t) \dot{\bf q}_t \geq 0 $ and $ \dot{ \bf q}^{\tau}_t T({\bf q}_t, t) \dot{\bf q}_t = 0 $, Eq.(1) leads to 
\begin{equation}
  \dot{ \bf q}_t \cdot \nabla \psi ({\bf q}_t, t) \geq 0 \; .
\end{equation}
This equation implies that the deterministic dynamics cannot decrease the fitness: The speed of state variable $\dot{ \bf q}_t$ is in the same direction of the gradient of the fitness function $\nabla \psi ({\bf q}_t, t)$. If the ascendant matrix is positive definite, i.e. $\dot{ \bf q}^{\tau}_t A({\bf q}_t, t) \dot{\bf q}_t > 0 $ for any nonzero $\dot{ \bf q}_t$, the fitness of the system always increases.
Hence, the first law clearly states that the system has the ability to find the local fitness peak determined by the initial condition.

I have one remark. From the mathematical theory of dynamical systems, there are in general three kinds of attractors \cite{guckenheimer}: point, periodic, and chaotic (strange). The point attractors have been well explored in biological life evolution since the work of Wright \cite{wright}, corresponding the fitness peaks. 
Other two kinds of attractors have also been observed in biology \cite{murray,may}. The content of the first law has been fully discussed in a recent review of evolutionary game dynamics in a different approach \cite{hofbauer}, whose connection to the first law will be clear in next section. I emphasize that there is no connection between different attractors in the first law.

The tendency implied in Eq.(1) to approach an attractor, represented by the nonnegative ascendant matrix which is similar to the friction in mechanics, and to remain there has been amply discussed by Aristotle. This law gives a reference point to define species and other relevant quantities in a clean manner, if stochasticity could be ignored.

\subsection{ Second Law }

The central question arises that how does one describe the evolutionary dynamics quantitatively and generally? 

I postulate that the dynamics of the system is governed by following special form of stochastic differential equation, which consists of four dynamical elements, the semi-positive definite symmetric ascendant matrix $A$, the anti-symmetric transverse matrix $T$, the scalar function called fitness function $\psi$, and the stochastic drive 
${\bf \xi}$:
\begin{equation}
  [ A({\bf q}_t, t) + T ({\bf q}_t, t) ] \dot{\bf q}_t 
     = \nabla \psi ({\bf q}_t, t) + {\bf \xi}({\bf q}_t, t) \; ,
\end{equation}
and supplemented by the following relationship:
\begin{equation}
  \langle {\bf \xi}({\bf q}_t, t) 
          {\bf \xi}^{\tau} ({\bf q}_{t'}, t') \rangle 
    = 2 A({\bf q}_t, t) \; \epsilon  \; \delta(t-t') \; ,
\end{equation}
and $\langle {\bf \xi}({\bf q}_t, t) \rangle = 0$.
The connection of these two equations to conventional approaches will be discussed in next section. 
In Eq.(4) I have assumed the stochastic drive is Gaussian with zero mean. Factor 2 is a convention choice for the present formulation, and $\epsilon$ is a positive numerical constant, which for many situations might be set to be unity, $\epsilon = 1$, without affecting the artificial life description. The relationship between the stochastic drive and the ascendant matrix expressed by Eq.(4) guarantees that the ascendant $ A({\bf q}_t, t)$ is semi-positive definite and symmetric.
The average $ \langle ... \rangle $ is carried over the dynamics of the stochastic drive, and $\delta(t)$ is the Dirac delta function.

The dynamical effect of the stochastic drive ${\bf \xi}({\bf q}_t, t)$
on fitness is random: It may either increase or decrease the fitness.
With above interpretation, the static effect natural selection is represented by the gradient of fitness function, 
$\nabla \psi({\bf q}_t, t)$. 
The clear and graphical discussion of such fitness function was first given by Wright in the discussion of biological life evolution \cite{wright}. The tempo of national selection is represented by the ascendant and transverse matrices. Eq.(3) states that the gradient of fitness, the stochastic drive, the ascendant dynamics, and the transverse dynamics, must be balanced to generate the evolution dynamics. 

The fitness function $\psi$ is similar to a potential: It is in fact opposite in sign to the typical potential energy used in physics and chemistry. If it is further independent of time and is bounded above, the stationary distribution function $\rho({\bf q}, t=\infty)$ for the state variable, the probability density to find the system at ${\bf q}$ in state space, is expected to be a Boltzmann-Gibbs distribution:
\begin{equation}
   \rho({\bf q}, t=\infty) = \frac{1}{Z} \exp \left\{ 
                    \frac{ \psi({\bf q}) }{ \epsilon } \right\} \; ,
\end{equation}
with ${Z} = \int d^n {\bf q} \exp \left\{ {\psi({\bf q}) }
/{ \epsilon } \right\} $ the partition function, the integration over whole state space.
Its justification will be given in next section. It is interesting to note that the dynamical aspects of evolution, the transverse and the ascendant matrices, do not explicitly show up in Eq.(5). 
This equation implies that the larger the constant $\epsilon$ is, the wider the equilibrium distribution would be, and more variation would be, or, the smaller the $\epsilon$ is, the narrower the distribution. In this sense $\epsilon$ may be called the evolution hotness constant.
The existence of such a Boltzmann-Gibbs type distribution suggests a global optimization. 

There are a few immediate and interesting conclusions to be drawn here. Near a fitness peak, say at ${\bf q}={\bf q}_{peak}$, I may expand the fitness function, $\psi({\bf q}) = \psi({\bf q}_{peak}) - ({\bf q} - {\bf q}_{peak})^{\tau} U ({\bf q} -{\bf q}_{peak})/2 
+ O(|{\bf q} -{\bf q}_{peak}|^3) $. 
Here $U$ is a positive definite symmetric matrix as a consequence at the fitness peak. The stationary probability density to find the system near this peak is of a typical Gaussian distribution:  
\begin{equation}
  \rho({\bf q}, t=\infty) \propto \;  \exp \left\{ - \frac{
   ({\bf q} - {\bf q}_{peak})^{\tau} U ({\bf q} - {\bf q}_{peak}) }
   {2\epsilon} \right\} \; .
\end{equation}
Thus, away from the fitness peak, the probability to find the system will be exponentially small. 

One may then wonder about how does the system move from one fitness peak to another? This process was first visualized by Wright \cite{wright}. The relevant mathematical calculation seems to be first done by Kramers \cite{kramers}, where it was shown that the stochastic drive must be involved. Quantitatively, the hopping from one peak to another must be aided by the stochastic drive. The dominant factor in the hopping rate $\Gamma$ is the difference in fitness between the peak and the highest point (saddle point ${\bf q}_{saddle}$) to cross the valley to another peak \cite{kramers,vankampen}: 
\begin{equation}
 \Gamma \propto \exp \left\{ -  
   \frac{\psi({\bf q}_{peak}) - \psi({\bf q}_{saddle}) }
        {\epsilon }\right\}\; .
\end{equation}
This rate can easily be exponentially small. It is a quantitative measure of robustness and stability. Hence it may explain the usual observation, for example, that species is rather stable if viewing the peak as a definition for species. Nevertheless, Eq.(7) grants the possibility to hop between peaks when the stochastic drive is finite.  

The second law clearly expresses that the evolutionary dynamics is probability in nature, has the ascendancy to large fitness in the adaptive landscape. 

\subsection{ Third Law }

The third law is a relationship law. It allows the definition of the connection of the current level of description to its lower and higher ones. It is a reflection of the hierarchical structure of the whole dynamics. The most important feature is that it acknowledges the existence of two time scales: micro and macro. 

Specifically, it may be stated as follows: The fitness function $\psi({\bf q},t)$ has the contribution from lower level in terms of time average on the time scale of current level, the contribution from the interaction among various components of the current level, and the contribution from higher level.
The stochastic drive $\xi({\bf q},t)$ is the remainder of all those contributions whose dynamics is fast on the time scale of current focus. Hence its average in time is zero. This stochastic contribution may be either unknown from a more fundamental level or unnecessary to be specified in details. Its probability distribution is approximated by a Gaussian distribution in the present article.
The stochastic drive determines the ascendant matrix $A({\bf q},t)$, and the transverse matrix $T({\bf q},t)$ should be further determined by the dynamics of the system.

The lower level contribution to fitness function $\psi({\bf q}, t)$ and stochastic drive $\xi({\bf q},t)$ may allow the computation of the intrinsic fitness landscape and the intrinsic source of evolution. However, this contribution tends to neglect the horizontal interaction among different components, which is usually nonlinear. On the other hand, the same and higher level contributions may suggest that a control mechanism, such as a feedback, may be from both of them in a large perspective. The combination of all three of them suggests that the evolution is nonlinear, asymmetric, mutually interactive, and stochastic, and may be controllable.

Such a hierarchical structure of artificial evolution has been extensively discussed \cite{simon}.   
There is, however, a degree of uncertainty and arbitrariness in the assignment of different levels of descriptions and the dichotomy of deterministic and stochastic terms in Eq.(3). This dilemma has been amply discussed in physical sciences \cite{vankampen}.  The present way to solve this problem will be proposed in next section in connection to usual dynamics. 

\subsection{ Fundamental Theorem of Evolution }

As implied in the first law, the ascendancy of the system is described by the ascendant matrix $A$, which in turn is completely determined by the stochastic drive according to the stochasticity-ascendancy relation, Eq.(4). The discussion followed Eq.(7) indicates that the ability of system to find a better fitness peak, no only the local fitness peak, or, to reach the global equilibrium, is guaranteed by the stochastic drive. This suggests that Eq.(4) is a statement on the unification of the two completely opposite tendencies: adaptation and randomization. 

A relation similar to Eq.(4) was also recognized long ago by Fisher in the biological life evolution \cite{fisher}, called fundamental theorem of natural selection. Eq.(4) may be called the fundamental theorem of evolution: It is independent of the fitness function but is the engine for evolution.

\section{ Conventional Formulation }

\subsection{ Standard Stochastic Differential Equation }

Now I make the connection between the dynamics described by Eq.(3) and (4) to the dynamical equations typically encountered in evolution.
I start with the standard stochastic differential equation: 
\begin{equation}
  \dot{\bf q}_{t} = {\bf f}({\bf q}_t, t) + 
                    {\bf \zeta}( {\bf q}_t , t) \; . 
\end{equation}    
Here ${\bf f}({\bf q}, t)$ is the deterministic nonlinear drive of the system, which includes effects from both other components and itself, and the stochastic drive is ${\bf \zeta}( {\bf q}, t)$, which differs from that in Eq.(3) but is governed by the same dynamics. 
For simplicity I will assume that ${\bf f}$ is a smooth function whenever needed. The importance and generality of such an equation has been known in evolution \cite{farmer2,hertz,rasmussen2} and its special limit of zero stochasticity in the context of evolutionary game dynamics was reviewed recently \cite{hofbauer}.

The stochastic drive in Eq.(8) is assumed to be Gaussian and white with the variance, 
\begin{equation}
 \langle {\bf \zeta}( {\bf q}_t , t) 
         {\bf \zeta}^{\tau}({\bf q}_{t'}, t') \rangle 
     = 2 D( {\bf q}_t, t) \; \epsilon \; \delta (t-t') ,
\end{equation}
and zero mean, $\langle {\bf \zeta}({\bf q}_t, t)\rangle = 0$. 
Again here $\langle ... \rangle$ indicates the average with respect to the dynamics of the stochastic drive. According to the physical science convention the semi-positive definite symmetric matrix $D=\{D_{ij}\}$ with $i,j=1,2, ..., n$ is the diffusion matrix.
Both the divergence and the skew matrix of the nonlinear drive ${\bf f}$ are in general non-zero:
\begin{equation}
 \nabla \cdot {\bf f} \neq 0, \; 
 \nabla \times {\bf f} \neq 0 \;.
\end{equation}
Here the divergence is explicitly $\nabla \cdot {\bf f} =  \sum_{j=1}^{n} \partial f_j /\partial q_j = tr(F)$, and the skew matrix $\nabla \times {\bf f}$ is twice the anti-symmetric part of the selection matrix $S$: $ ( \nabla \times {\bf f} )_{ij} = S_{ji} - S_{ij}$ with $ S_{ij} = \partial f_i /\partial q_j \ , \; i,j  =1,2, ..., n$.
The non-zero of the divergence leads to that the state space volume is not conserved: Ascendancy is implied. 
The non-zero of the skew matrix, or the asymmetry of the selection matrix $S$, implies the existence of the transverse matrix $T$. 

Now, I give an explicit construction which demonstrates the existence and uniqueness connection between Eqs. (3,4) and Eqs. (8,9). Assuming that both Eq.(3) and (8) describe the same dynamics. The speed $\dot{\bf q}_t $ is then the same in both equations. The connection from Eq.(3) to (8) is straightforward: 
Multiplying both sides of Eq.(3) by $ [A({\bf q}_t, t)  + T({\bf q}_t, t) ]^{-1} $ leads to Eq.(8).
The procedure from Eq.(8) to (3) is mathematically more involved.
 
Using Eq.(8) to eliminate the speed in Eq.(3), and noticing that the dynamics of noise and the state variable behave independently, I have 
\begin{equation}
  [A({\bf q}, t) + T({\bf q}, t)] {\bf f}({\bf q}, t) 
   =  \nabla \psi({\bf q}, t) \; ,
\end{equation}
and 
\begin{equation}
   [A({\bf q}, t) + T({\bf q}, t)] {\bf \zeta}({\bf q}, t) 
       = \xi({\bf q}, t) \; . 
\end{equation}
Here I have dropped the subscript $t$ for the state variable, because time $t$ is now a parameter.
Those two equations suggest a rotation in state space.

Multiplying Eq.(12) by its transpose on each side and carrying out the average over stochastic drive, I have
\begin{equation}
 [A({\bf q}, t) + T({\bf q}, t) ] D({\bf q}, t) 
 [A({\bf q}, t) - T({\bf q}, t) ] = A({\bf q}, t) \; .
\end{equation}
In obtaining Eq.(13) I have also used Eq.(4) and (9).
Eq.(13) suggests a duality between the standard stochastic differential equations and Eq.(3): A large ascendant matrix implies a small diffusion matrix. It is a generalization of the Einstein relation for $T=0$ case \cite{einstein}.

Next I define an auxiliary matrix function 
\begin{equation}
   G({\bf q}, t) = [A({\bf q}, t) + T({\bf q}, t) ]^{-1} \; .
\end{equation}
Here the inversion `${-1}$' is with respect to the matrix. Using the 
property of the fitness function $\psi$: 
$\nabla \times \nabla \psi = 0$ [$(\nabla \times \nabla \psi)_{ij} = (\nabla_i \nabla_j -\nabla_i \nabla_j)\psi $ ], 
Eq.(11) leads to
\begin{equation}
  \nabla \times [ G^{-1} {\bf f}({\bf q}) ] = 0 \; ,
\end{equation}
which gives $n(n-1)/2$ conditions.
The generalized Einstein relation, Eq.(13), leads to the following 
equation
\begin{equation}
   G + G^{\tau} = 2 D \; ,
\end{equation}
which readily determines the symmetric part of the auxiliary matrix 
$G$, another $n(n+1)/$ conditions. The auxiliary function may be formally solved as an iteration in gradient expansion:
\begin{equation}
  G = D + Q \; , 
\end{equation}
with $Q = \lim_{j \rightarrow \infty} \Delta G_j $, 
$\Delta G_j = \sum_{l=1}^{\infty} (-1)^l [ (S^{\tau})^l 
 \tilde{D}_j S^{-l} + (S^{\tau})^{-l} \tilde{D}_j S^l ] $,
$\tilde{D}_0 = DS - S^{\tau}D$, 
$\tilde{D}_{j \geq 1} = ( D + \Delta G_{j-1} ) 
     \left\{ [\nabla \times (D^{-1} + \Delta G_{j-1}^{-1} ) ] 
     {\bf f} \right\} ( D - \Delta G_{j-1} )$. 
At each step of solving for $\Delta G_j$ only linear algebraic equation is involved. One can verify that the matrix $Q$ is anti-symmetric. For a simple case a formal solution of such algebraic equation was given in \cite{ao2002}, and an explicitly procedure was found for generic cases in \cite{kat}. Eq.(17) is a result of local approximation: If the selction matrix $S$, the diffusion matrix $D$ are constant in space, the exact solution only contains the lowest order contribution in gradient expansion: $Q= \Delta G_j = \Delta G_0 $. I regard Eq.(17) as the artificial life solution to Eq.(15) and (16), because it preserves all the fixed points of deterministic drive ${\bf f}$. 
The connection from Eq.(8) to (3) is therefore uniquely determined: 
\begin{equation}
  \left\{ \begin{array}{lll}
  \psi({\bf q}, t) & = & \int_C d{\bf q}' \cdot 
          [ G^{-1}({\bf q}') {\bf f}({\bf q}') ] \\
  A({\bf q}, t) & = & [G^{-1}({\bf q})+(G^{\tau} )^{-1}({\bf q})]/2 \\
  T({\bf q}, t) & = & [G^{-1}({\bf q})-(G^{\tau} )^{-1}({\bf q})]/2
  \end{array} 
     \right. \; . 
\end{equation} 
Here the sufficient condition $\det(A+T)\neq 0$ is used, and the end and initial points of the integration contour $C$ are $\bf q$ and ${\bf q}_0$ respectively. The construction of fitness function and other quantities from the conventional approach, summarized in Eq.(18), appears to be given for the first time by the present author.

The always existence of the fitness function and the associated ascendant matrix and transverse matrix may be understood in the follow way. After all, in the evolution of Artificial Life, a selection must be made. It could be purposely, or, determined by the available condition. If there would be no selection, that is, everything would have an equal probability, there would be no evolution. 
The fitness function is simply a reflection of this fact, and the no selection case is a constant fitness function everywhere in the phase space. No structure would be expected in this case. 
The stochasticity links to ascendant matrix by the stochascitiy-ascendance relation or the fundamental theorem of evolution. It is a remarkable unification on the two apparent opposite tendencies. 
However, there are situations that a selection would not affect the total fitness. The transverse matrix reflects this dynamical conservation of fitness.

I should point out that in the absence of stochastic drive, i.e., $\epsilon = 0$ in Eq.(4) and (9), above connection remains unchanged.

\subsection{ Fokker-Planck Equation }

In many experimental studies in biological life evolution, a question is often asked on the distribution of the state variable as a function of time instead of focusing on the individual trajectory of the system. This implies that either there is an ensemble of identical systems or repetitive experiments are carried out. To describe this situation, I need a dynamical equation for the distribution function in the phase space. This goal can be accomplished by the so-called Fokker-Planck equation, or the difussion equation \cite{vankampen}. 

In this subsection, another procedure to find the equation for distribution function is presented. It is natural from a theoretical physics point. This procedure will establish that the fitness function $\psi$ in Eq.(1) indeed plays the role of potential energy in the manner envisioned by Wright, and the steady state distribution will be indeed given by Eq.(3). The present starting point will be the second law, Eq.(1), not the standard stochastic differential equation, Eq.(8), from which most previous derivations started.

The existence of both the deterministic and the stochastic drives in Eq.(3) suggests that there are two well separated time scales in the system: the microscopic or fine time scale to describe the stochastic drive and the macroscopic or course time scale to describe the system motion. The former time scale is much smaller than the latter. This separation of time scales further suggests that the macroscopic motion of the system has an "inertial": it cannot response instantaneously to the microscopic motion. To capture this feature, I introduce a small constant inertial "mass" $m$ and a kinetic momentum vector ${\bf p}$ for the system. The state space is then enlarged: It is now a $2n$-dimensional space. The dynamical equation for the system takes the form:
\begin{equation}
   \dot{\bf q}_t = {\bf p}_t /m \; , 
\end{equation}
which defines the kinetic momentum, and 
\begin{equation}
    \dot{\bf p}_t = - [A({\bf q}_t, t) + T({\bf q}_t, t) ] {\bf p}_t /m 
                         + \nabla \psi({\bf q}_t, t) 
                         + {\bf \xi}({\bf q}_t, t) \; ,
\end{equation} 
which is the extension of Eq.(3). I note that there is no dependent of ascendant matrix $A$ and the stochastic drive on the kinetic momentum ${\bf p}$. The Fokker-Planck equation in this enlarged state space can be immediately obtained \cite{vankampen}:
\begin{equation}
 \left\{ \partial_t + \frac{\bf p}{m} \cdot \nabla_{\bf q} 
           + \overline{\bf f} \cdot \nabla_{\bf p} 
  - \nabla_{\bf p}^{\tau} A \left[\frac{\bf p}{m} + \nabla_{\bf p} 
\right]
\right\} 
   \rho({\bf q}, {\bf p}, t) = 0 \; . 
\end{equation}
Here $\overline{\bf f} = {\bf p}^{\tau} T /m + \nabla_{\bf q} \psi $, and $t$, 
${\bf q}$, and ${\bf p}$ are independent variables. 
The subscripts in the $\partial$ and $\nabla$ indicate the differentiation with respect to indicated variable only. 
The stationary distribution can be found, when the fitness function is time-independent and bounded above, as \cite{vankampen} 
\begin{equation}
 \rho({\bf q},{\bf p},t=\infty) = \frac{1}{\cal Z} \exp\left\{-\frac{ 
   {{\bf p}^2 }/{2m} - \psi({\bf q}) }{\epsilon } \right\} \; ,
\end{equation}
with ${\cal Z} = \int d^n{\bf q} d^n {\bf p} \exp\{ - [ {\bf p}^2 /2m 
- \psi({\bf q}) ]/\epsilon \}$ the partition function. 
There is an explicit separation of state variable and its kinetic momentum in Eq.(22). The elimination of the momentum in the small mass limit will not affect this distribution. Hence, Eq.(22) confirms that the expected Boltzmann-Gibbs distribution, Eq.(5) from the Eq.(3) and (4), is the right choice. 

I proceed to outline the procedure to find the Fokker-Planck equation corresponding to Eq.(3) and (4) without the kinetic momentum ${\bf p}$. I first illustrate how to recover Eq.(3) from Eq.(19) and (20). 
In the limit of $m \rightarrow 0$, the fast dynamics of kinetic momentum ${\bf p}_t$ can always follows the motion of slow dynamics of state variable ${\bf q}_t$. Hence I may set $\dot{\bf p}_t = 0 $ in Eq.(20) and replace the kinetic momentum using Eq.(19), which is then Eq.(3) after moving the speed to the left-side of equation. 
For the Fokker-Planck equation, the explicit separation of the kinetic momentum and state variable in the stationary distribution gives the guidance on the procedure: The resulting Fokker-Planck equation must be able to reproduce this feature. The Fokker-Planck equation is then found as
\begin{equation}
  \partial_t \rho({\bf q},t) 
   = \nabla^{\tau} [ - {\bf f}({\bf q}) - \Delta{\bf f}({\bf q}) 
                     + D({\bf q}) \nabla ] \rho({\bf q}, t) \; ,
\end{equation}
with $\Delta {\bf f}$ the solution of the equation 
$ \nabla \cdot \Delta {\bf f} + \Delta {\bf f}\cdot \nabla \psi 
  - \nabla \cdot [ G T G^{\tau} \nabla \psi ] = 0$.
If the probability current density is defined as ${\bf j}({\bf q},t) 
\equiv ( {\bf f} + \Delta{\bf f} - D \nabla ) \rho({\bf q},t)$, the 
Fokker-Planck equation is a statement of the probability continuity: 
\begin{equation}
  \partial_t \rho({\bf q},t) + \nabla \cdot {\bf j}({\bf q},t) = 0 \; .
\end{equation} 
The stationary state corresponds to the condition 
 $ \nabla \cdot {\bf j}({\bf q}, t=\infty) = 0 $.
One may verify that the stationary distribution $\rho({\bf q}, t=\infty)$ in Eq.(3) is indeed the time independent solution of the Fokker-Planck equation: The stationary probability current 
\begin{equation}
 {\bf j}({\bf q}, t=\infty) = (G T G^{\tau} + \Delta {\bf f}) \nabla \psi({\bf q}) \; \rho({\bf q},t=\infty) \; ,
\end{equation}
and $\nabla \cdot {\bf j}({\bf q},t=\infty) = 0 $.

\subsection{ Detailed Balance Condition }

There is an important class of evolution dynamics in which the anti-symmetric matrix $Q = 0$. Under this condition, the transverse matrix $T=0$, and  $\Delta {\bf f} = 0$. The Fokker-Planck equation becomes
\begin{equation}
  \partial_t \rho({\bf q},t) 
   = \nabla^{\tau} [ - {\bf f}({\bf q})) 
                     + D({\bf q}) \nabla ] \rho({\bf q}, t) \; ,
\end{equation}
and the stationary probability current is everywhere zero in state phase:
\begin{equation}
   {\bf j}({\bf q}, t=\infty) = 0 \; .
\end{equation}
In this situation one may find that
\begin{equation}
   \nabla \psi({\bf q})= D^{-1}({\bf q}) {\bf f}({\bf q}) \; ,
\end{equation}
and $A = D^{-1} $.
The fitness function and the connection between Eq.(3) and the standard stochastic differential equation Eq.(8) can be directly read out from equations. This is the well-known symmetric dynamics in physical sciences \cite{vankampen}. This zero probability current condition is usually called the detailed balance condition.  

\section{ Examples }

In this section I discuss two examples. The first one is of predator-prey model like. In this model I illustrate how to approximately compute the fitness function $\psi$, the ascendant matrix $A$ and the transverse matrix $T$, that is, how to make the connection between the conventional formulation and Eq.(3) and (4). Second example is a current model in biological life evolution. It will be demonstrated that it can be discussed within the present mathematical formulation. 

\subsection{ Predator-Prey Model }
   
The example whose dynamical equation is in the form of standard stochastic differential equation is the generic predator-prey process. Under the diffusion approximation, both the diffusion matrix $D$ and the deterministic drive can be obtained from the master equation. The diffusion approximation is valid when a large number of birth and death events occurs on the macroscopic time scale \cite{vankampen}.

I now give an explicit demonstration of how to obtain Eq.(3) from Eq.(8) for a two component case. Here $q_1$ and $q_2$ represent numbers of two species in a habitat. 
I assume the spatial distribution is uniform. The deterministic drive ${\bf f}$ consists of two positive terms, birth and death:
\begin{equation}
    f_i({\bf q}) = f_{ib}({\bf q}) - f_{id}({\bf q}) \; \; i=1,2 \; ,
\end{equation}
with the subscripts $b$ and $d$ stand for the birth and death respectively. Under the diffusion approximation, the stochastic drive is \cite{vankampen}
\begin{equation}
 \zeta_i({\bf q},t) = \sqrt{ f_{ip}({\bf q}) } \zeta_{ip}(t) +
                      \sqrt{ f_{id}({\bf q}) } \zeta_{id}(t)
                         \; \; i=1,2 \; ,
\end{equation}
with $\zeta_{ip}(t), \zeta_{id}(t)$ are unity random variables and possible correlation among them. Therefore the diffusion matrix $D$ can be readily obtained, which is what needed below. I remark that the equation similar to predator-prey equation has been emerged in the study of the robustness of the gene regulatory network of phage $\lambda$ \cite{zhu}.

The construction of Eq.(3) from Eq.(8) will be given to the lowest order in the gradient expansion. The usefulness of this approximated construction can be illustrated for following two reasons. First, in many practical applications, lowest order approximation is already enough \cite{zhu}, because it is exact in the strictly linear case. Second, several salient features of the connection becomes apparent without undue mathematical complications. An important quantity is the selection matrix $S$. According to the definition following Eq.(13),
\begin{equation}
    S_{11} = \nabla_1 f_1 \, ,    S_{12} = \nabla_2 f_1 \, ,
    S_{21} = \nabla_1 f_2 \, ,    S_{22} = \nabla_2 f_2  \; .
\end{equation}
Eq.(10) will not change under the gradient approximation. In the lowest order gradient approximation, Eq.(9) becomes simple. I collect them here: 
\begin{equation}
  \left\{ \begin{array}{lll}
   G S^{\tau} - S G^{\tau} & = & 0  \\
   G          + G^{\tau}   & = & D 
  \end{array} 
     \right. \; . 
\end{equation} 
In two dimensions the matrix manipulation is particularly straightforward. I note that any $2\times 2$ matrix $M$ can uniquely decomposed in terms of Pauli matrices, $\sigma_i$ with $i=1,2,3$, and the identical matrix ${\bf 1}$:
\[
  M = M_1 \sigma_1 + M_2 \sigma_2 + M_3 \sigma_3 
      + tr(M)/2 \; {\bf 1} \; ,
\]
with $tr$ denotes the trace and $\sigma_1 = \left( \begin{array}{ll}
                 0 & 1  \\
                 1 & 0 
    \end{array} \right) $, 
    $\sigma_2 = \left( \begin{array}{ll}
                 0 & -i  \\
                 i & 0 
    \end{array} \right) $, 
    $\sigma_1 = \left( \begin{array}{ll}
                 1 & 0  \\
                 0 & -1 
    \end{array} \right) $, and here $i = \sqrt{-1}$.  
Using this relationship, the equation for antisymmetric part of the auxiliary matrix $G = D + Q$ from Eq.(32) is
\begin{equation}
  Q S^{\tau} + S Q =  (S D - D S^{\tau}) \; .
\end{equation} 
Using the matrix decomposition and the properties of Pauli matrices, 
I obtain
\begin{equation}
  Q = { (S D - D S^{\tau}) }/ { tr(S) } \; .
\end{equation} 
Note that for the $2\times2$ matrix $M$
\[
   \left( \begin{array}{ll}
                 M_{11} & M_{12} \\
                 M_{21} & M_{22}  
    \end{array} \right) ^{-1} = \frac{1}{ \det(M) }
    \left( \begin{array}{ll}
                 M_{22} & - M_{12} \\
               - M_{21} & M_{11}  
    \end{array} \right) 
\]
The ascendant matrix $A$ and the transverse matrix $T$ can be found according to Eq.(8):
\begin{equation}
  \left\{ \begin{array}{lll}
    \psi({\bf q}) & = & \int_C d{\bf q}' \cdot 
          [ G^{-1}({\bf q}') {\bf f}({\bf q}') ] \\
    A({\bf q}) & = &
         \left( \begin{array}{ll}
                 D_{22} & - D_{12} \\
               - D_{12} & D_{11}  
         \end{array} \right)  /\det(G)  \\
    T({\bf q}) & = & {-Q}/{\det(G)} 
  \end{array} 
     \right. \; . 
\end{equation} 
In two dimensions, $\det(G) = \det(D) + \det(Q)=D_{22}D_{11}-D_{12}^2 + Q_{12}^2 $ and is obviously non-negative.

To summarize, the computation of quantities in Eq.(3) from Eq.(8) is as follows: First, to establish Eq.(8) from the artificial life problem, continuous approximation should be used. If the problem is given in terms of master equation, the usual diffusion approximation will be employed, which gives both the diffusion matrix and the deterministic drive \cite{vankampen}. After this is done, the procedure prescribed in section III.A is employed to find the ascendant matrix, the transverse matrix and the fitness matrix. According to the artificial life problem, an additional approximation may be used to reduce computation effort but with the desired accuracy, as done here as well as in Zhu {\it et al.} \cite{zhu} for a biological life evolution problem.
 
\subsection{ Symmetry-Breaking Model }

The symmetric evolution dynamics was explicitly discussed by Stewart for speciation \cite{stewart}:
\begin{equation}
  \dot{\bf q }_t = \nabla \phi({\bf q}_t ) \; .
\end{equation}
This is a deterministic equation. The need for stochastic modeling was also mentioned by Stewart. 

In the light of present discussion, the diffusion matrix $D$ is equivalent to a diagonal matrix. Hence the fitness function can be directly obtained: $\psi({\bf q}_t ) = \phi({\bf q}_t )$.
The steady state distribution is then given by the Boltzmann-Gibbs like distribution, Eq.(3). All the statistical physics methodology can then be applied here. Naturally one may make use of the idea of symmetry-breaking as a way for self-organization needed for speciation. I refer readers to the beautiful discussion presented by Stewart \cite{stewart}.  
One may even make use of the self-consistent mean-field approximation, a powerful mathematical tool in statistical physics \cite{adami,goldenfeld}, to search for the indication of symmetry-breaking.

\section{ Discussions }

Before further going to further discussion on the implication of the present mathematical formulation, it should be kept in mind that above three laws must be regarded as what the evolution dynamics might be. They are by not means the exact description.
Having made this statement, I nevertheless remark that although proposed three laws for evolution dynamics are based on the continuous approximation, it is possible that main features discussed in the present article survive in discrete cases.

One may ask why to use Eq.(3) and (4) instead of more conventional Eq.(8) and (9): After all their equivalence has been demonstrated above. Here I offer three reasons to favor Eq.(3) and (4):
 
1) Quantities presented in Eq.(3) can be directly related to experimental observation. For example, Eq.(7) gives a direct connection between the fitness function and the population density in steady state. By observing the dynamical behaviors, information on the ascendant and transverse matrices can be obtained. 
Also, Eq.(7) can relate stability to the fitness function. This direct contact with experimental data is an indication of the autonomy of the focused level of description.

2) Eq.(8) and (9) lack the visualizing ability for the global dynamics behavior. For example, in a nonlinear dynamics with multiple local maxima, it is not clear from Eq.(8) and (9) which maximum is the largest one, and how easy it might be to mover from one maximum to another.
One could find this answer by a direct real time calculation. 
But this is usually computationally demanding, if not impossible.

3) Eq.(3) and (4) give an alternative modeling of evolutionary dynamics, which can be advantageous in certain situations. For example, the direct use of fitness function in Stewart's modeling \cite{stewart} makes the symmetry-breaking idea very transparent from statistical physics' point of view.  

One may argue that given the importance of stochasticity, the present Gaussian white noise assumption may not be general enough. This is certainly true. For a critique of this sort the present article already serves its purpose: It is a starting point.
An example is the apparent more noisy genetic algorithm \cite{holland,channon}. This is one of most successful algorithms to model adaptation and stochasticity. At a first glance it may not appear to fit into present formation: It is difficult to define fitness function and stochasticity appears more wild than that expressed by Gaussian white noise. However, the very fact that there is a selection means a fitness function must exist. A coarse grain average in time, that is, over suitable large number of generations, a continuous approximation in time is still possible. This has been well demonstrated in the source of genetic algorithm: The diffusion approximation in population genetics of biological life evolution has been demonstrated to be useful. 

Finally, I point out a unique computational advance of the present formulation. For a system has many stable fixed points, it is computationally very expensive, if possible at all, to find the best stable fixed point from the conventional approach represented by Eq.(8). It is already difficult enough to compare the relevant stability of two stable fixed points according to Eq.(8). 
For example, it is not possible to compare which Nash equilibrium is better in the conventional approach. This comparison is normally not discussed \cite{hofbauer}.
On the other hand, the fitness function in (3) directly provides a graphical solution to this question: the larger the fitness the better.  Furthermore, the difference of the fitness peak to the nearby saddle point gives a direct measurement of the global stability, as expressed by Eq.(7). 
No a real time simulation is needed to find the global evolution trend: Given conventional equation Eq.(8), the fitness function can be constructed according to Eq.(18). This offers an alternative routine for artificial life evolution, certainly a tremendous computational edge.
 
\section{ Conclusions }

In the present article I have postulated three laws to describe the evolutionary dynamics of artificial life with four dynamical elements. The most fundamental equation, the second law, has been expressed in a unique form of stochastic differential equation. 
The fitness function and stochasticity have been emphasized in the present formulation. 
I have demonstrated that present laws are consistent with more conventional approaches, but appear more suitable to discuss stability and other phenomena quantitatively. The fitness function is precisely defined and is viewed as the standard to make selection. The stochasticity is viewed as the source of creativity, and its effect is formulated in a transparent relation. 
The present formulation offers an alternative routine to determine the global evolution trend in addition to direct numerical simulation.

{\ }

\noindent{\bf Acknowledgement:} 
This work was supported in part by a USA NIH grant under HG002894-01.
 
{\ }


\begin{thebibliography}{99}

\bibitem{open}
 Bedau, M. A {\it et al.} (2000) Open problems in artificial life.
  Artificial Life 6: 363-376.
\bibitem{farmer1}
 Farmer, J.D. and Packard, N.H. (1986) Evolution, game, and learning: 
   models for adaption in machines and nature. Physica D 22: vii-xii. 
\bibitem{rasmussen}
 Rasmussen, S. {\it et al.} (2001) Ansatz for dynamical hierarchies. 
  Artificial Life. 7: 329-353.
\bibitem{al2}
 Artificial Life II (1992) edited by C.G. Langton, C. Taylor, 
   J.D. Farmer, and S. Rasmussen, Addison-Wesley, Redwood City.
\bibitem{simon}
 Simon, H.A. (1996) The Sciences of the artificial, 3rd edition.
   MIT Press, Cambridge.
\bibitem{etxeberria}
 Etxeberria, A. (2000) Artificial evolution: creativity and possible.  
   in Artificial Life VII, edited by M.A. Bedau, J.S. McCaskill, 
   N.H. Packard, and S. Rasmussen, MIT Press, Cambridge: 555-562.
\bibitem{jones}
 Jones, S. (2003) Organizing relations and emergences. In Artificial
   Life VIII, edited by R.K. Standish, M.A. Bedau, and H.A. Abbass, 
   MIT Press, Cambridge: 418-422.         
\bibitem{vankampen}
 Kampen, van N.G. (1992) Stochastic processes in physics and
  Chemistry. Elsevier, Amsterdam. 
\bibitem{adami}
 Adami, C. (1998) Introduction to artificial life. Springer-Verlag,
  Berlin.
\bibitem{farmer2}
 Farmer, J.D. (1990) A Rosetta stone for connectionism.  Physica D 42:
   153-187.
\bibitem{hofbauer}
 Hofbauer, J. and Sigmund, K. (2003) Evolutionary game dynamics.
   Bull. Am. Math. Soc. 40: 479-519.
\bibitem{hertz}
 Hertz, J. (2003) Computing with attractors. In The Handbook of Brain 
   Theory and Neural Networks, 2nd edition, edited by M.A. Arbib: 
   248-252.       
\bibitem{channon}
 Channon, A.D. and Damper, R.J. (1998) Perpetuating evolutionary
   emergence. in From Animals to Animates 5, edited by R. Pfeifer, 
   B. Blumberg, J.-A. Meyer and S.W. Wilson, MIT Press, Cambridge: 
\bibitem{maynardsmith}
 Maynard Smith, J. (1982) Evolution and the theory of games. Cambridge 
  University Press, Cambridge. 
\bibitem{stewart}
 Steward, I. (2003) Self-organization in evolution: a mathematical  
  perspective. Phil. Trans. R. Soc. Lond. A361: 1101-1123.
\bibitem{guckenheimer}
 Guckenheimer, J. and Holmes, P. (1997) Nonlinear oscillations,
  dynamical systems, and bifurcations of vector fields.
  Springer-Verlag, Berlin.
\bibitem{wright}
 Wright, S. (1932) The roles of mutation, inbreeding, crossbreeding 
  and selection in evolution. Proceedings of the Sixth International
  Congress of Genetics, 1: 356-366. 
\bibitem{murray}
 Murray, J.D. (2002) Mathematical biology. v. 1. Springer, New York.
\bibitem{may}
 May, R.M. (1981) (ed) Theoretical ecology: principles and
   applications. Second edition. Blackwell Scientific, Oxford.
\bibitem{kramers}
 Kramers, H.A. (1940). Brownian motion in a field of force and the 
  diffusion model of chemical reactions. Physica 7: 284-304.
\bibitem{fisher}
 Fisher, R.A. (1930) The genetical theory of natural selection,
  Clarendon, Oxford.
\bibitem{rasmussen2}
 Rasmussen, S. and Barrett, C.L. (1995) Elements of a theory of
   simulation. In Advances in Artificial Life, edited by F. Moran, 
A. Moreno, J.J. Morelo, and P. Chacon, Springer, Berlin: 515-529.
\bibitem{einstein} 
 Einstein, A. (1905) Ann. Physik 17: 549-56.
\bibitem{ao2002}
 Ao, P. (2002) Stochastic force defined evolution in dynamical 
  systems. (http://it.arXiv.org/find/physics/1/Ao/0/1/0/past/3/0)
\bibitem{kat}
 Kwon, C., Ao, P. and Thouless, D. J. (2003). Structure of Stochastic
  dynamics near fixed points 
 (submitted to PNAS, available upon request).
\bibitem{zhu}
 X.-M. Zhu, L. Yin,  L. Hood, and P. Ao, Calculating biological
   behaviors of epigenetic states in phage $\lambda$ life cycle,
   Functional and Integrative Genomics (2004) 
   (DOI:  10.1007/s10142-003-0095-5 ).
\bibitem{goldenfeld}
 Goldenfeld, N. (1992) Lectures on phase transitions and the
   renormalization group. Addison-Wesley, Reading.
\bibitem{holland}
  Holland, J.H. (1998) Emergence: from chaos to order. Addison-Wesley,
  Reading.               

\end{thebibliography}
\end{document}